\documentclass[reprint,superscriptaddress,amsmath,amssymb,longbibliography,twocolumn]{revtex4-2}

\usepackage{amsmath}
\usepackage{mathtools}
\usepackage{amssymb}
\usepackage{graphicx}
\usepackage{lipsum}
\usepackage{bm}
\usepackage[colorlinks, linkcolor=blue, citecolor=blue, urlcolor=blue, breaklinks=true]{hyperref}
\usepackage{color,dsfont,multirow,booktabs}
\usepackage{braket}
\usepackage{tabularx}
\usepackage{lipsum,booktabs}
\usepackage[normalem]{ulem}
\usepackage{ulem}
\usepackage{subfiles}
\usepackage{subcaption}
\usepackage{caption}
\usepackage{ragged2e}
\usepackage{orcidlink}
\usepackage{txfonts,times}

\newcommand\av[1]{\langle #1\rangle}
\newcommand\dg[1]{#1^{\dagger}}
\newcommand\numberthis{\addtocounter{equation}{1}\tag{\theequation}}

\begin{document}
\date{\today}

\title{Super-Optimal Charging of Quantum Batteries via Reservoir Engineering}

\author{Borhan Ahmadi\orcidlink{0000-0002-2787-9321}}
\email{borhan.ahmadi@ug.edu.pl}
\address{International Centre for Theory of Quantum Technologies, University of Gdansk, Jana Bażyńskiego 1A, 80-309 Gdansk, Poland}
\author{Pawe{\l} Mazurek\orcidlink{0000-0003-4251-3253}}
\email{pawel.mazurek@ug.edu.pl}
\address{Institute of Informatics, Faculty of Mathematics, Physics and Informatics, University of Gdańsk, Wita Stwosza 63, 80-308 Gdańsk, Poland}
\author{Shabir Barzanjeh}
\address{Department of Physics and Astronomy, University of Calgary, Calgary, AB T2N 1N4 Canada}
\author{Pawe{\l} Horodecki}
\address{International Centre for Theory of Quantum Technologies, University of Gdansk, Jana Bażyńskiego 1A, 80-309 Gdansk, Poland}

\begin{abstract}

Energy dissipation, typically considered an undesirable process, has recently been shown to be harnessed as a resource to optimize the performance of a quantum battery. Following this perspective, we introduce a novel technique of charging in which coherent charger-battery interaction is replaced by a dissipative interaction via an engineered shared reservoir. We demonstrate that exploiting collective effects of the engineered shared reservoir allows for extra optimization giving rise to optimal redistribution of energy, which leads to a significant enhancement in the efficiency of the charging process. The article unveils the intricacies of built-in detuning within the context of a shared environment, offering a deeper understanding of the charging mechanisms involved. These findings apply naturally to quantum circuit battery  architectures, suggesting the feasibility of efficient energy storage in these systems. Moreover, the super-optimal charging offers a practical justification for charger-battery configurations.

\end{abstract}

\maketitle

\textit{Introduction}.---The role of energy transfer is of primary importance in industry. In recent years, significant research has been devoted to examining how to efficiently facilitate energy transfer between two quantum systems by means of a direct interaction between the systems in various contexts such as quantum heat engines, quantum thermal logic gates, quantum thermal transistors, and quantum batteries \cite{PhysRevLett.99.177208,PhysRevLett.116.200601,PhysRevE.109.064146,RevModPhys.96.031001,PhysRevLett.132.260403,PhysRevLett.120.117702,PhysRevLett.122.047702}. Quantum batteries \cite{PhysRevE.87.042123} represent a cutting-edge advancement in the field of energy transfer and storage, leveraging the principles of quantum mechanics to potentially revolutionize how we catch and store energy in quantum devices \cite{PhysRevE.87.042123,PhysRevLett.120.117702,PhysRevLett.122.047702,PhysRevResearch.5.013155,PhysRevLett.120.117702,Rodríguez_2024,PhysRevA.107.032218,kamin2023steady,kamin2020non,binder2015quantacell,PhysRevA.107.042419,RevModPhys.96.031001}. Unlike traditional batteries, quantum batteries utilize quantum phenomena such as entanglement \cite{PhysRevLett.120.117702,PhysRevLett.122.047702,PhysRevResearch.5.013155, PhysRevLett.120.117702} to facilitate energy storage processes that could surpass the limitations of conventional systems. Other approaches such as using quantum optimal control \cite{Rodríguez_2024,PhysRevA.107.032218}, quantum catalysis processes \cite{PhysRevA.107.042419}, superabsorption \cite{quach2022superabsorption}, phase transition \cite{barra2022quantum} and indefinite causal order \cite{PhysRevLett.131.240401} can also be used to improve the efficiency of the charging process, where the charging time decreases as the battery's capacity increases, a stark contrast to classical batteries where larger capacities typically mean longer charging times.

While these contributions have significantly advanced our understanding of quantum batteries, the search for enhanced efficiency drives us to explore new avenues for improving energy storage, most notably boosting its energetic efficiency and power. A novel approach to optimize the efficiency of quantum battery charging leverages reservoir engineering to induce nonreciprocity \cite{PhysRevLett.77.4728, PhysRevLett.112.133904, PhysRevX.5.021025, Toth2017, Barzanjeh2017, PhysRevApplied.4.034002, PhysRevX.7.041043, 2303.04358, PhysRevLett.120.060601} by which, as recently shown, energy transfer to the battery may be significantly boosted \cite{PhysRevLett.132.210402}. Reservoir engineering, a crucial technique in quantum technology, involves the deliberate design and manipulation of a system's interaction with its environment to achieve specific quantum states or dynamics. This method has found applications across various fields, including quantum information processing \cite{PhysRevLett.77.4728,verstraete2009quantum}, quantum simulation \cite{barreiro2011open}, and precision measurement \cite{murch2013observing}. By tailoring the environmental interactions, researchers can counteract decoherence and dissipation, traditionally seen as detrimental, and instead stabilize desired quantum states or induce coherent dynamics \cite{PhysRevLett.107.080503}.

In this work, we identify a novel scenario in which, unlike in Ref. \cite{PhysRevLett.132.210402}, nonreciprocity condition is not satisfied, and yet the sheer presence of an engineered shared reservoir allows for significant enhancement of efficiency of the quantum battery. We first demonstrate that the optimal charging for any charging process is done by optimisation over characteristics of the external energy source for a given charging device which is determined by a built-in detuning of the charger-battery structure. We then show that the engineered shared reservoir allows for \textit{further} optimization of the charging process over the internal structure of the charging device (parameters of the coupling with the engineered reservoir), resulting in \textit{super-optimal} charging. We provide the microscopic description of this intriguing phenomenon through the master equation formalism. We show that energy dissipation, typically considered a damping and an undesirable process, can be harnessed as a resource to further optimize the performance of a quantum battery. Specifically, replacing the coherent (direct) interaction between the charger and the battery with a dissipative (indirect) interaction enables a significant increase in transfer of energy to the battery. Generally, creating coherent interaction is more challenging than designing broadband lossy systems. At last, we show that super-optimal charging allows for the energy of the battery in the charger-battery setting to exceed that of the battery directly coupled to the external laser (no charger mediating the process), providing further justification for charger-battery configurations \cite{farina2019charger}. This research opens up possibilities for exploring chiral and topological properties in systems utilizing dissipative coupling.

\begin{figure}[t]
\centering
\includegraphics[width=8.5cm]{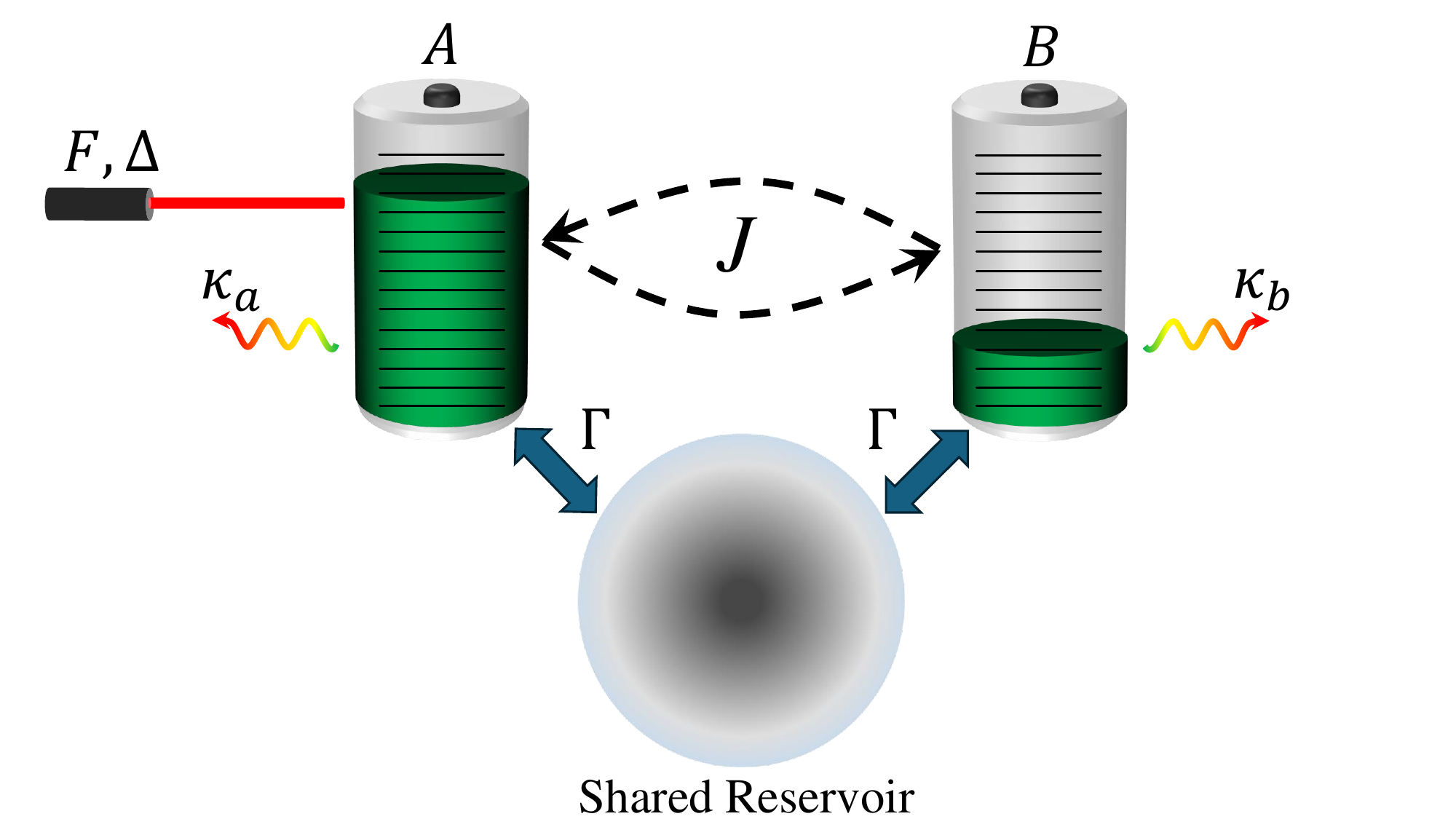}
    \captionsetup{justification=justified}
    \caption{Schematic representation of the charging process. A quantum charger $A$ coherently interacts with a quantum battery $B$ with a coupling rate $J$. Both the charger and the battery decay into a shared reservoir at the decay rate $\Gamma$ establishing an effective dissipative interaction between them. A laser field with amplitude $F$ is utilized to supply energy to the charger. Here, $\kappa_a$ and $\kappa_b$ describe the local damping rates of each mode.\justifying}
    \label{Scheme}
\end{figure}

\textit{Quantum Battery Model}.---Figure \ref{Scheme} illustrates the schematic of the system in a charger-battery configuration \cite{farina2019charger}, including a harmonic oscillator serving as the charger with a resonance frequency of $\omega_a$ and a local damping rate $\kappa_a$. This charger interacts with another mode acting as the battery, characterized by a frequency of $\omega_b$ and a damping rate $\kappa_b$. The coherent interaction between the charger and the battery is described by $J=|J|e^{i\phi}$. An external classical drive with a frequency $\omega_L$ and amplitude $F$ provides energy to the charger during the charging process. 

The Hamiltonian describing this driven bipartite system $AB$ can be written, in the rotating frame with respect to $\omega_L$, as:
\begin{equation}\label{Hamiltonian0}
H = H_{AB} + H_L.
\end{equation}
The Hamiltonian $H_{AB}$ is given by $H_{AB} = H_a + H_b + H_{int}$, where $H_a = \Delta a^\dagger a$ ($\hbar=1$) and $H_b = \Delta b^\dagger b$ represent the free Hamiltonians of the charger and the battery, respectively. Here, $a$ ($b$) and $a^\dagger$ ($b^\dagger$) are the annihilation and creation operators of the charger (battery), and $\Delta = \omega_L - \omega$ denotes the local detuning of the input laser beam from the optical charger resonance frequency. For simplicity, we assume that the charger and the battery are in resonance, i.e., $\omega_a = \omega_b = \omega$. 
The interaction Hamiltonian, $H_{int}$, is described as $J a^\dagger b + J^*ab^\dagger$, and the term $H_L = F(e^{i\Delta t} a + e^{-i\Delta t} a^\dagger)$ accounts for the laser field applied to the charger for feeding energy into the system (under rotating wave approximation \cite{breuer2002theory}).
As we construct our charging system $AB$, a \textit{built-in} eigen-frequencies of the Hamiltonian $H_{AB}$ are established $\omega_{\pm} = \omega \pm |J|$ (see Appendix \ref{AppendixA}). This in turn creates a built-in detuning of 
\begin{align}\label{in}
    \Delta_{in} = \pm |J|,
\end{align} where the subscript $in$ indicates its built-in origin. This suggests that in a \textit{non-dissipative} charging process, to optimize energy transfer to the battery the detuning of the external field $\Delta$ be equal to this built-in detuning $\Delta_{in}$. This phenomenon will be rigorously proved in the following sections.

In addition, we also consider a dissipative interaction between the charger and the battery, mediated by their mutual coupling to a shared reservoir characterized by a coupling rate of $\Gamma$. This interaction is realized by adiabatically eliminating the reservoir's degrees of freedom. From a practical perspective, the shared reservoirs could be an additional damped cavity or a waveguide \cite{PhysRevX.5.021025,PhysRevLett.120.140404}. The dynamics of the system can be described by Lindblad master equation \cite{PhysRevX.5.021025}
\begin{equation}\label{ME}
\dot{\rho}=-i[H,\rho] + \Gamma\mathcal{D}_z[\rho] + \kappa_a\mathcal{D}_a[\rho] + \kappa_b\mathcal{D}_b[\rho],
\end{equation}
where $\mathcal{D}_O[\cdot]=O\cdot O^\dagger-\frac{1}{2}\{O^\dagger O,\cdot\}$ represents the dissipative super-operator resulting from the coupling to the reservoirs. Here $\kappa_{a(b)}$ is the local decay rate of the charger (battery) and $z=p_a a+p_b b$, with $p_a$ ($p_b$) denoting the coupling strength between the charger (battery) with the shared reservoir with decay rate $\Gamma$. Note that the second term in Eq. \eqref{ME} represents the dissipation of both modes into the shared reservoir, resulting in dissipative coupling between the two modes. The last two terms account for local damping of each of the modes due to their coupling ports or the surrounding environment \cite{PhysRevX.5.021025,chang2011slowing}. As we are going to explore, in presence of dissipation the optimal detuning is not anymore the built-in detuning, but includes also corrections depending on the type of the dissipation. 

We define the internal energy of each subsystem as $E_i= \omega\text{tr}{\big[\rho_i i^\dagger i}\big]$\label{energy} with $i=a,b$ \cite{alicki1979quantum}. Given the assumption of zero temperature for all reservoirs, the states of subsystems remain pure if the initial state of the whole system is a pure product state \cite{kossakowski1972quantum}. This in turn implies that all the internal energy of the battery can be extracted as ergotropy \cite{Allahverdyan_2004}. Therefore, in the following, we will only discuss the internal energies of the battery and the charger.

\textit{Charging Dynamics}.---We now analyze the local dynamics of the system using the master equation \eqref{ME}. The evolution of the first and second moments is given by
\begin{align}
    \frac{d\av{a}}{dt} &= -\frac{\Gamma_{a} + \kappa_{a} + 2i\Delta}{2} \av{a} 
    - i\left(J+i\mu\frac{\Gamma}{2}\right)\av{b} - iF \label{L1}, \\
    \frac{d\av{b}}{dt} &= -\frac{\Gamma_{b}+\kappa_{b} + 2i\Delta}{2}\av{b}
    -i\left(J^*+i\mu^*\frac{\Gamma}{2}\right)\av{a} \label{L2}, \\
    \frac{d\av{\dg{a}a}}{dt} &= -(\Gamma_{a}+\kappa_{a})\av{\dg{a}a}
    -2\Re\left\{i\left(J+i\mu\frac{\Gamma}{2}\right)\av{\dg{a}b}\right\} 
    - 2F\Im\left\{\av{a}\right\} \label{L3}, \\
    \frac{d\av{\dg{b}b}}{dt} &= -\left(\Gamma_{b}+\kappa_{b}\right)\av{\dg{b}b}
    +2\Re\left\{i\left(J-i\mu\frac{\Gamma}{2}\right)\av{\dg{a}b}\right\} \label{L4}, \\
    \frac{d\av{\dg{a}b}}{dt} &= -\frac{\Gamma_{a}+\kappa_{a}+\Gamma_{b}+\kappa_{b}}{2}\av{\dg{a}b} 
    - i\left(J^*+i\mu^*\frac{\Gamma}{2}\right)\av{\dg{a}a} \nonumber, \\
    &\quad + i\left(J^*-i\mu^*\frac{\Gamma}{2}\right)\av{\dg{b}b} + iF\av{b}, \label{L5}
\end{align}
where $\mu=-p_{b}p^{*}_{a}$ and $\Gamma_i=\Gamma|p_{i}|^{2}\label{p}$ with $i=a,b$. The symbols $\Re$ and $\Im$ indicate the real and imaginary parts, respectively. As long as both $p_{a}$ and $p_{b}$ are non-zero, we re-scale them such that $|\mu| = 1$, and the remaining factor is absorbed into $\Gamma$.
    
\begin{figure}[t]
    \centering
    \includegraphics[width=0.5\textwidth]{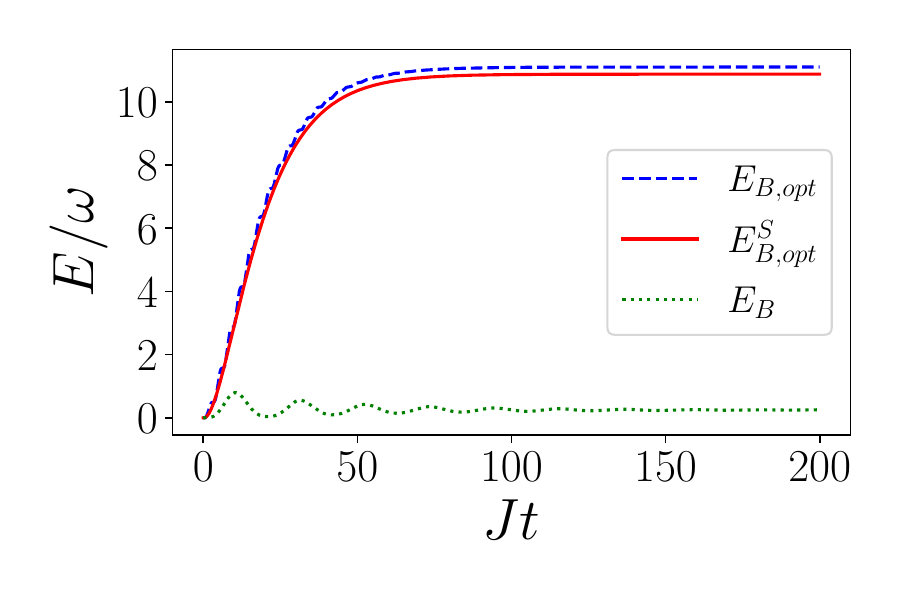}
    \captionsetup{justification=justified}
    \caption{Optimal energy of the battery in the conventional procedure (dashed blue line) and the shared reservoir-assisted procedure (solid red line) versus $Jt$. Here we have $|J|=0.2$, $\kappa_a=0.05$, $\kappa_b=0.01$ and for the shared reservoir-assisted procedure $\Gamma=0.4$, $p_a=p_b=1$, $\phi=0, \pi$. In both procedures we selected the corresponding optimal detuning $\Delta_{opt}$. The green solid line is for conventional procedure when detuning is not optimal, for instance here $\Delta=0$, to emphasize the huge effect of optimal detuning on the charging process. \justifying}
    \label{EnergyB}
\end{figure}

\textit{Optimal Charging}.---From the set of differential equations of motion \eqref{L1}-\eqref{L5}, we calculate the energy of the charger and the battery in the steady state 
\begin{equation}\label{EnergyA0}
    E_A = \frac{4F^2\mathcal{K}_{bb}}{|\mathcal{J}(i\Gamma,\phi) - \mathcal{K}_{ab}|^2},
\end{equation}
\begin{equation}\label{EnergyB0}
    E_B=\frac{4F^2\mathcal{J}(\Gamma,\alpha)}{|\mathcal{J}(i\Gamma,\phi) - \mathcal{K}_{ab}|^2},
\end{equation}
where $\mathcal{K}_{bb}=4\Delta^2 + (\Gamma_b + \kappa_b)^2$, $\mathcal{K}_{ab}=\left(2\Delta + i(\Gamma_a + \kappa_a)\right)\left(2\Delta + i(\Gamma_b + \kappa_b)\right)$, $\mathcal{J}(x,\theta)=4|J|^2 + x^2 - 4|J|x\cos{\theta}$ and $\alpha+\phi=\pi/2$. From Eqs. \eqref{L1}-\eqref{L5} we can see that $\mathcal{J}(x,\theta)$ effectively incorporates the interactions facilitating energy transfer between the charger and the battery, while $\mathcal{K}_{ij}$ describes processes restricting energy transfers such as energy dissipation or detuning. We first start the analysis by examining the conventional quantum battery scenario (the absence of the shared reservoir). After setting $\Gamma=0$ in Eqs. \eqref{EnergyA0} and \eqref{EnergyB0}, the steady-state energies of the charger and the battery are, respectively, determined as
\begin{equation}\label{EnergyA1}
    E_A = \frac{4F^2\mathcal{K}_{bb}^{\Gamma=0}}{|\mathcal{J}(i\Gamma,\phi) - \mathcal{K}_{ab}|_{\Gamma=0}^2},
\end{equation}
\begin{equation}\label{EnergyB1}
    E_B = \frac{16F^2\mathcal{J}(0,\alpha)}{|\mathcal{J}(i\Gamma,\phi) - \mathcal{K}_{ab}|_{\Gamma=0}^2}.
\end{equation}
We choose the initial states of both charger and battery to be ground states. For the ideal charging process where there exists no local dissipations, i.e. $\kappa_a=\kappa_b=0$, by optimizing $E_{B}$ over $\Delta$ we find that the optimal detuning equals to the build-in detuning (\ref{in}): $\Delta_{opt}=\Delta_{in}$. In fact, in the case $\Delta=\Delta_{in}$ we have $\mathcal{J}(i\Gamma,\phi)=\mathcal{K}_{ab}$, which gives rise to infinite energy transfer to the battery. This phenomena of infinite energy transfer indicates the crucial significance of the built-in detuning $\Delta_{in}$. However, in realistic charging processes, local dissipations are inevitable. Therefore, for nonzero local dissipation rates $\kappa_{a(b)}$, optimizing $E_{B}$ over $\Delta$ the optimal charging process results in
\begin{equation}\label{Delta}
    \Delta_{opt}=\pm\sqrt{\Delta^2_{in} - \frac{\kappa^2}{8}},
\end{equation}
where $\kappa^2=\kappa_a^2 + \kappa_b^2$. Note that in this case, infinite energy transfer to the battery is not feasible. This is because the denominator in Eq. \eqref{EnergyB1} takes the form $|4J^2 - (2\Delta + i\kappa_a)(2\Delta + i\kappa_b)|^2$, which is always greater than zero due to the imaginary terms resulting from local dissipation rates $\kappa_{a(b)}$. In Fig. \ref{EnergyB} we plot the time evolution of $E_B(t)$ for the optimal charging procedure (dashed blue line). Additionally, in Fig. \ref{EnergyB} we illustrate $E_B(t)$ for a non-optimal detuning case (green solid line) by taking $\Delta=0$, to underscore the substantial influence of optimal detuning on charging process. It is worth mentioning that Eq. \eqref{Delta} implies the important fact that the charging laser filed comprehensively perceives not solely the coherent interaction $J$, but also energy dumping to the local baths. The interactions with the baths show their effects via the dissipation rates $\kappa_{a(b)}$. Therefore, in presence of local dissipations, $\kappa$ is in fact the correction to the built-in detuning $\Delta_{in}$ (a fact overlooked in our earlier work \cite{PhysRevA.107.042419}). In other words, the optimal detuning is influenced by the entirety of the interactions. We also notice that in the conventional scenario $\Delta_{opt}$ depends only on the strength of the coherent interaction, and not on its phase $\phi$.

We now move to the crucial question of the paper: can the presence of an engineered shared reservoir enhance the optimal conventional charging process?
Let us first consider adding the engineered shared reservoir to the charging process by setting  $\Gamma\neq 0$, and optimizing $E_B$ in the stationary limit.
In this scenario we derive the optimal detuning to be again close to the built-in detuning $\Delta_{in}$, with  more convoluted corrections due to interactions with both local and shared reservoirs (see Appendix \ref{AppendixA}). In this case, $\Delta_{opt}$ depends also on the interaction phase $\phi$. As illustrated in Fig. \ref{EnergyB}, adding the shared reservoir does not provide a straightforward advantage over the optimal conventional charging process.

Yet, there is another way to exploit the engineered shared reservoir, which we outline below.

\textit{Super-Optimal Charging}.---Here we demonstrate that if instead of {\it adding} a shared reservoir to the charging process, as discussed above, we {\it replace} the coherent interaction by the shared reservoir, then the resulting dissipative interaction between the charger and the battery will enable us to {\it significantly outperform} the optimal conventional charging process. To better understand the mechanism, we first provide the insight into the microscopic dynamics underlying the concept of super-optimal charging by examining the master equation \eqref{ME} and then introduce our approach aimed at surpassing the optimal conventional charging strategy discussed in the previous section. The master equation \eqref{ME} may be recast as follows
\begin{align*}\label{ME3}
    \dot{\rho} &= -i[H,\rho] + (\Gamma_a+\kappa_a)\mathcal{L}_a[\rho] + (\Gamma_b+\kappa_b)\mathcal{L}_b[\rho] \\
    &+ \Gamma\mathcal{L}_a^b[\rho] + \Gamma\mathcal{L}_b^a[\rho],\numberthis
\end{align*}
where
\begin{equation}\label{Lxy}
   \mathcal{L}_x^y[\cdot] =: x\cdot y^\dagger - \frac{1}{2}\{x^\dagger y,\cdot\}.
\end{equation}
Eq. \eqref{ME3} implies that the shared reservoir dissipator $\mathcal{D}_z[\rho]$ from (\ref{ME}) in practice affects the charging process in two distinct manners: firstly, via $\Gamma_x\mathcal{D}_x[\rho]$ it causes energy dissipation of the charger and the battery into the shared reservoir at rates $\Gamma_a$ and $\Gamma_b$, respectively. Secondly, via $\Gamma\mathcal{D}^y_x[\rho]$ it facilitates energy exchange between the charger and the battery at rate $\Gamma$. Therefore, $\Gamma\mathcal{D}^y_x[\rho]$ provides a dissipative interaction between the charger and the battery paving the way for more transfer of energy. Notably, even with the direct coherent coupling turned off ($J=0$), energy exchange still remains feasible through the shared reservoir with rate $\Gamma$. This observation leads to an intriguing phenomena in the realm of charging quantum batteries, opening up new possibilities for energy transfer manipulation, as discussed in the following. Note that in quantum optics the same generator of the form \eqref{ME3} is also used to explain the phenomenon of collective resonance fluorescence \cite{RPuri}.

Motivated by this, we explore a distinct scenario where the direct coherent interaction is replaced by a dissipative interaction provided by a shared reservoir (as a result, no built-in detuning anymore). This means that in the equations of motion \eqref{L1}-\eqref{L5} we put $J=0$ and $\Gamma\neq0$. Therefore in this scenario the laser field exclusively interacts with the charger, leading to resonance at $\omega_L=\omega$. The pivotal question arises: with the direct coherent interaction replaced by a shared reservoir with the dissipation rate $\Gamma$, what value of $\Gamma$ enables us to either to recover or to surpass the optimal result obtained in the conventional procedure? Intuitively, considering the equations of motion \eqref{L1}-\eqref{L5}, since $J$ is always accompanied with $i\Gamma/2$, the answer lies in selecting $\Gamma=2|J|$, as confirmed in the following.

\begin{figure*}[ht]
    \centering
    \begin{subfigure}[b]{0.45\textwidth}
        \includegraphics[width=\textwidth]{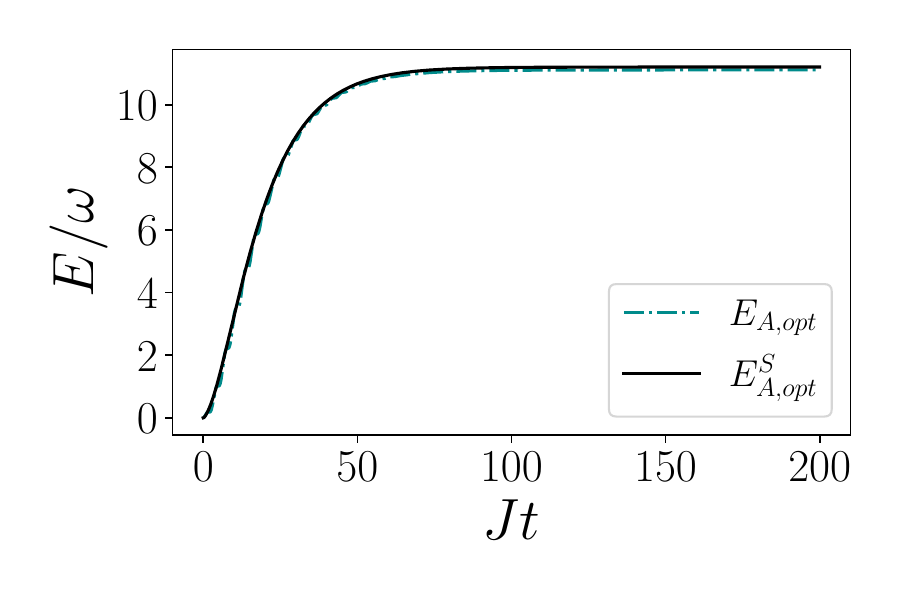}
        \caption{}
        \label{EnergyASb}
    \end{subfigure}
    \hspace{0.01\textwidth} 
    \begin{subfigure}[b]{0.45\textwidth}
        \includegraphics[width=\textwidth]{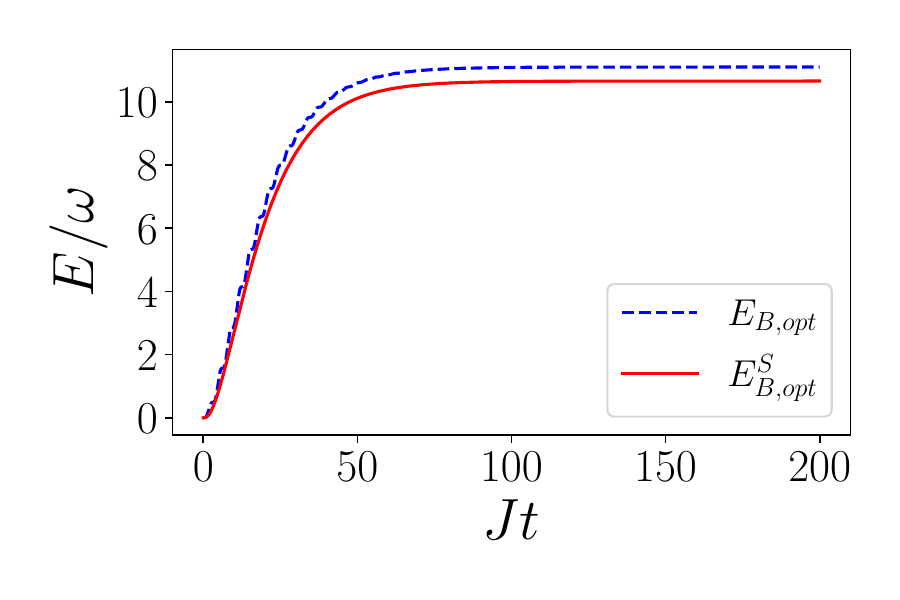}
        \caption{}
        \label{EnergyBSb}
    \end{subfigure}
    \captionsetup{justification=justified}
    \caption{$(\textbf{a})$ Energies of the charger $E_A$ and $(\textbf{b})$ the battery $E_B$ versus $Jt$ for the optimal conventional scenario and the shared reservoir-assisted scenario. In the conventional scenario we have $F=0.1$, and $|J|=0.2$. In the shared reservoir-assisted scenario $F=0.1$, $J=0$, $\Gamma=0.4$, $p_{a}=1$, $p_{b}=1$. In both procedures we have $\kappa_a=0.05$, $\kappa_b=0.01$ and the corresponding optimal detuning $\Delta_{opt}$. \justifying}
    \label{EnergyABop}
\end{figure*}

Replacing the coherent interaction $J$ by the shared reservoir with dissipation rate $\Gamma$ yields the steady state energies of the charger and the battery, respectively, as
\begin{equation}\label{EnergyA20}
    E^S_A = \frac{4 F^2\mathcal{K}_{bb}}{|\mathcal{J}(i\Gamma,\phi) - \mathcal{K}_{ab}|_{J=0}^2},
\end{equation}
\begin{equation}\label{EnergyB20}
    E^S_B = \frac{4 F^2\mathcal{J}_{J=0}(\Gamma,\alpha)}{|\mathcal{J}(i\Gamma,\phi) - \mathcal{K}_{ab}|_{J=0}^2},
\end{equation}
where superscript $S$ denotes the presence of the shared reservoir while no coherent interaction on. From Eqs. \eqref{EnergyA20} and \eqref{EnergyB20} it can be shown that the optimal detuning here is zero, $\Delta_{opt}=0$ (see also Eq. (16) of Appendix \ref{AppendixA}). This aligns with physical intuition, as deactivating the coherent interaction confines the laser field influence solely to the charger, disregarding the remainder of the system. Thus, by substituting $\Delta=0$ into Eqs. \eqref{EnergyA20} and \eqref{EnergyB20}, we arrive at
\begin{equation}\label{EnergyA2}
    E^S_{A, opt} = \frac{4F^2\mathcal{K}_{bb}^{\Delta=0}}{|\mathcal{J}(i\Gamma,\phi) - \mathcal{K}_{ab}|_{\Delta,J=0}^2},
\end{equation}
\begin{equation}\label{EnergyB2}
    E^S_{B, opt} = \frac{4F^2\mathcal{J}_{J=0}(\Gamma,\alpha)}{|\mathcal{J}(i\Gamma,\phi) - \mathcal{K}_{ab}|_{\Delta,J=0}^2},
\end{equation}
\begin{equation}\label{xis}
    \xi^S_{opt} = \frac{4F^2(\mathcal{J}\left(\Gamma,\theta)+\mathcal{K}_{bb}\right)_{\Delta,J=0}}{|\mathcal{J}(i\Gamma,\phi) - \mathcal{K}_{ab}|_{\Delta,J=0}^2},
\end{equation}
where $\xi\equiv E_A+E_B$ denotes the total energy within the entire system. In Fig. \ref{EnergyABop}, we present a plot depicting the battery energy for $\Gamma=0.4$, $J=0$ and $\Delta_{opt}=0$, comparing it with the optimal conventional scenario in which $|J|=0.2$. Again it is seen that no benefits are provided by the shared reservoir. We should stress that in Fig. \ref{EnergyABop} and the following figures energy is plotted against $Jt$ where the scaling factor $J$ is taken from the conventional scenario. This choice in fact ensures a fair comparison across both scenarios as long as we set $\Gamma$ in the shared reservoir-assisted scenario to be twice the value of $|J|$ in the conventional scenario. This is because in the equations of motion \eqref{L1}-\eqref{L5}, $J$ always comes with $i\Gamma/2$.
\begin{figure*}[ht]
    \centering
        \begin{subfigure}[b]{0.45\textwidth}
\includegraphics[width=\textwidth]{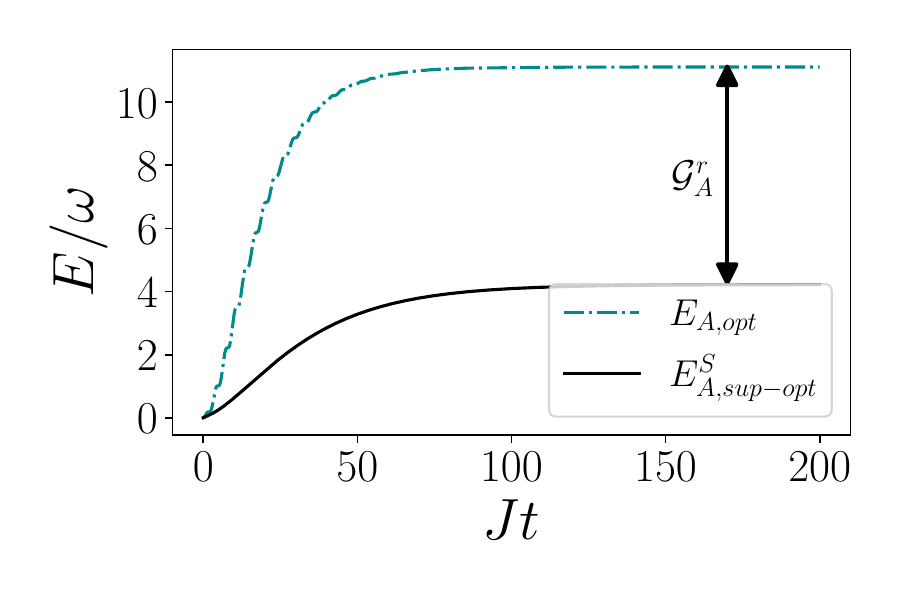}
        \caption{}\label{EnergyASc}
    \end{subfigure}
    \hspace{0.01\textwidth} 
    \begin{subfigure}[b]{0.45\textwidth}
\includegraphics[width=\textwidth]{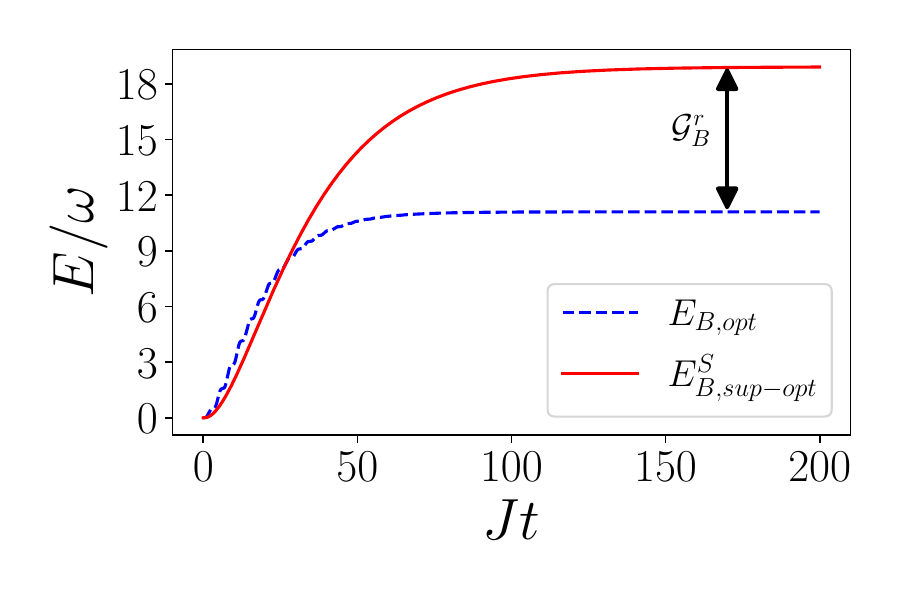}
        \caption{}\label{EnergyBSc}
    \end{subfigure}
    \captionsetup{justification=justified}
    \caption{$(\textbf{a})$ Energies of the charger $E_A$ and $(\textbf{b})$ the battery $E_B$ versus $Jt$ for the optimal conventional scenario and the shared reservoir-assisted scenario. The quantity $\mathcal{G}_{A(B)}^r$ quantifies the redistribution gap caused by the engineered shared reservoir. For the conventional scenario we have $F=0.1$, and $|J|=0.2$. For the shared reservoir-assisted scenario $F=0.1$, $J=0$, $\Gamma=0.4$, $p_{a}=(\kappa_a/\kappa_b)^{1/4}$, $p_{b}=(\kappa_a/\kappa_b)^{-1/4}$. In both procedures we chose $\kappa_a=0.05$, $\kappa_b=0.01$ and the corresponding optimal detuning $\Delta_{opt}$. \justifying}
    \label{EnergyABso}
\end{figure*}
However, as we show below, leveraging the structure brought by  the shared reservoir, we are able to optimize the process by manipulating relative values of $\Gamma_{a(b)}$. Examining equations of motion we see that changing $\Gamma_a$ and $\Gamma_b$ by altering $p_a$ and $p_b$ can impact the energy of the charger and the battery. Let us denote by $y$ the ratio $y=\frac{|p_{a}|^2}{|p_{b}|^2}$, we then have $\Gamma_a/\Gamma_b=y$. Now rewriting Eqs. \eqref{EnergyA2}-\eqref{xis} as functions of $x$, we find that both the total energy of the setup and the energy of the battery are optimized for
\begin{align}\label{x}
    y=\sqrt{\frac{\kappa_a}{\kappa_b}},
\end{align}
while energy of the charger reaches maximum in the limit $y\rightarrow 0$. We see that the essence of optimization (\ref{x}) is to properly reflect imbalance between local dissipation strengths $\kappa_{a(b)}$ in the ratio of $p_{a(b)}$. Consequently, we will be talking about a \textit{super-optimal} charging process in a scenario with shared reservoir optimized in this way. The corresponding super-optimal stationary energies are given by: 
\begin{equation}\label{EnergyA2opt}
    E^S_{A, sup-opt} = \frac{4F^2}{\kappa_a^2},
\end{equation}
\begin{align}\label{EnergyB2opt}
    E^S_{B, sup-opt} = \frac{4\Gamma^2F^2}{\kappa_a \kappa_b (2 \Gamma +\sqrt{\kappa_a} \sqrt{\kappa_b})^2},
\end{align}
\begin{equation}\label{xisopt}
    \xi^S_{sup-opt} = \frac{4F^2(\kappa_a \kappa_b^2 + 2\Gamma\kappa_b \sqrt{\kappa_a\kappa_{b}} + \Gamma^2(\kappa_a + \kappa_b))}{\kappa_a^2\kappa_b(2\Gamma + \sqrt{\kappa_a\kappa_b})^2}.
\end{equation}
As a direct result of the optimization, we observe a significant increase in the total energy of the system $\xi^S$ (see Appendix \ref{AppendixB}). This indicates that the optimization helps preserve more energy in the charger-battery system against the dissipation. Furthermore, the second intriguing result, as detailed below, is that optimization not only enhances the total energy within the entire system but allows for channelling, in the most efficient way, the energy into the battery $B$ as well.

As shown in Fig. \ref{EnergyASc} applying further optimization, through manipulation of $\Gamma_{a(b)}$ on the charging process, causes $E_{A, sup-opt}^S$ to be substantially lower than $E_{A, opt}$, emphasizing again the significant benefits of having an engineered shared reservoir in the charging process. At the same time, as shown in Fig. \ref{EnergyBSc}, we observe that the super-optimized energy of the battery $E_{B, sup-opt}^S$ (solid red line) significantly surpasses the optimal conventional energy $E_{B, opt}$. This further justifies the name \textit{super-optimal} charging process, since it outperforms the optimal conventional charging process. Comparing Fig. \ref{EnergyASb} with Fig. \ref{EnergyASc} and Fig. \ref{EnergyBSb} with Fig. \ref{EnergyBSc} the remarkable phenomena that we observe is the redistribution of total energy within the entire system due to the optimization, with a notable increase of energy in the desired location—namely, the battery $B$.

As illustrated in Fig. \ref{EnergyABso}, to quantify the redistribution gap caused by the engineered shared reservoir, we also define the quantity $\mathcal{G}_{A(B)}^r\equiv E^S_{A(B), sup-opt} - E_{A(B), opt}$. As we are interested in the energy of the battery, we focus on
\begin{align*}
    \mathcal{G}_{B}^r &= 4 \Gamma ^2 F^2 \left(\frac{4}{(\kappa_{a}+\kappa_{b})^2 \left((\kappa_{a}-\kappa_{b})^2-4 \Gamma ^2\right)}+
    \frac{1}{\kappa_{a} \kappa_{b}
   \left(2 \Gamma +\sqrt{\kappa_{a} \kappa_{b}}\right)^2}\right) 
\end{align*}
and observe that in settings characterized by significant difference in local dissipation rates $\kappa_{a}$ and $\kappa_{b}$, this function is positive. This attests for charging scenarios in which we observe a guaranteed advantage of shared reservoir over coherent couplings.
\begin{figure}[t]
    \centering
    \includegraphics[width=0.5\textwidth]{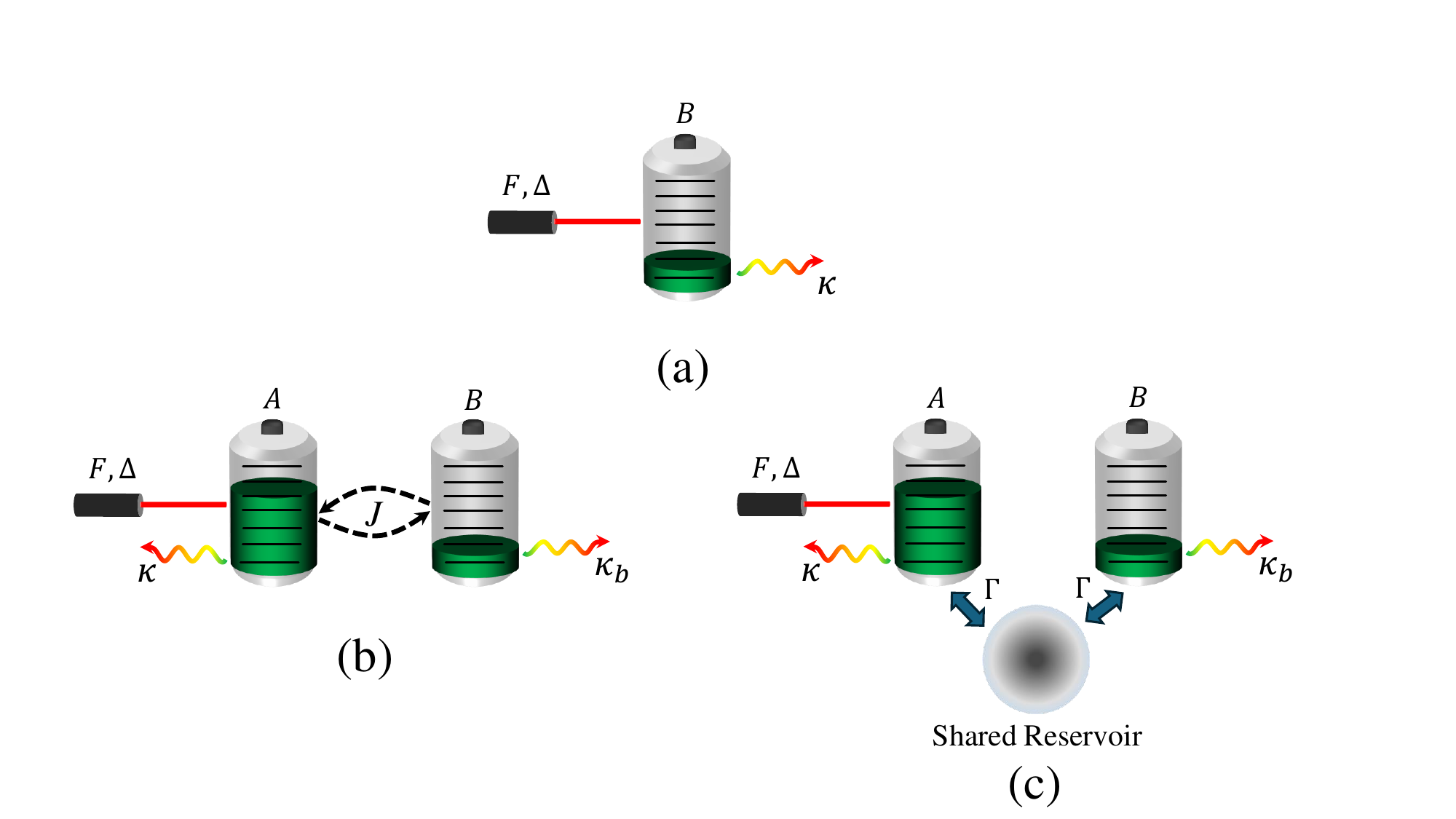}
    \captionsetup{justification=justified}
    \caption{Three charging scenarios: (\textbf{a}) direct coupling of the battery to the laser, (\textbf{b}) charger-mediated setting, with charger coherently coupled to the battery, and (\textbf{c}) charger-mediated setting, with charger dissipatively coupled to the battery. In (\textbf{b}) and (\textbf{c}) the presence of system $A$ (charger) isolates the battery and assures smaller dissipation $\kappa_b<\kappa$. \justifying}
    \label{Schemeabc}
\end{figure}

\textit{In defense of charger-battery setups}.---
In fact, this advantage is significant enough to address the motivational issues behind charger-battery charging settings. Namely, Ref. \cite{farina2019charger} considered a separation of the charging setup into a charger, whose role is to receive energy from the laser drive, and battery, whose role is to store the energy, coupled to the charger. The battery, while coupled coherently to the noisy charger, in experimental implementations can be much more isolated and therefore experiencing much smaller energy dissipation. However, in Ref. \cite{gangwar2024coherently} the Authors pointed to the lack of operational benefit of using this extended setting, compared to the simple model of a battery interacting directly with the laser field.

This problem, and the role of super-optimal charging in solving it, is illustrated in Fig. \ref{Schemeabc}. Three charging scenarios are presented there: (a) through a direct coupling of the battery to the laser, (b) in a charger-mediated setting with the charger coherently coupled to the battery, and (c) in a charger-mediated setting with the charger dissipatively coupled to the battery. We ask the following question: can we obtain higher energies in the stationary state of the battery, if we include a charger to intermediate between the laser and the battery? In this procedure, we keep the dissipation rate $\kappa$ of the party interacting with laser constant, and allow for smaller values of the dissipation rate $\kappa_b$ of the battery in the extended scenario, to account for its better isolation from the surroundings.

\begin{figure}[t]
    \centering
    \includegraphics[width=0.5\textwidth]{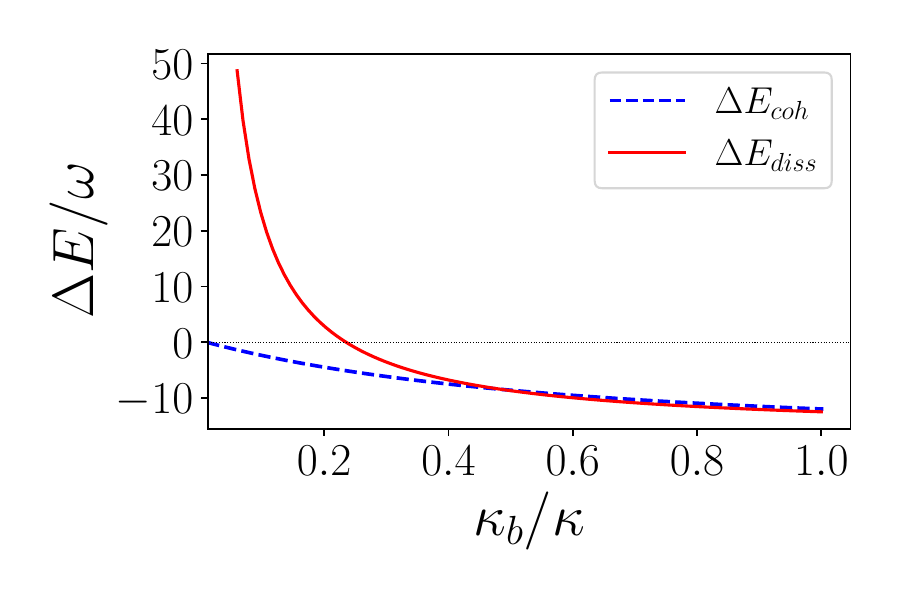}
    \captionsetup{justification=justified}
    \caption{Advantage in stationary energy of the battery $B$ in extended scenarios (corresponding to Fig. \ref{Schemeabc} (b) and (c)), compared to single-battery scenario (Fig. \ref{Schemeabc}(a)), expressed through $\Delta E_{coh}$ and $\Delta E_{diss}$, respectively,  for various dissipation rates $\kappa_b$. For scenario (a) we choose $F=0.1$, for (b) we have $F=0.1$, $J=0.2$, $\Gamma=0$, and for (c) we select $F=0.1$, $J=0$, $\Gamma=0.4$, and optimized values $p_{a}=(\kappa_a/\kappa_b)^{1/4}$, $p_{b}=(\kappa_a/\kappa_b)^{-1/4}$. In all scenarios we choose $\kappa=0.05$ and the corresponding optimal detuning $\Delta_{opt}$. We have $\lim_{\kappa_b\rightarrow 0}\Delta E_{diss}=\infty$ and $\Delta E_{coh}(0)=0.016$. \justifying}
    \label{DeltaE}
\end{figure}

We compare stationary energies achieved with extended scenarios (b) $E_B$ (given by Eqs. (\ref{EnergyB1}) and (\ref{Delta})) and (c) $E^S_{B, sup-opt}$ (given by Eq. (\ref{EnergyB2opt})) with the stationary energy of the scenario (a):
\begin{equation}\label{DE10}
    E_{single-batt}= \frac{4F^2}{\kappa_a^2},
\end{equation}
and obtain
\begin{align}\nonumber
\Delta E_{coh} & \equiv E_{B, opt} - E_{\text{single-batt}} \\
& = 4 F^2 \left( -\frac{4 J^2}{(\kappa + \kappa_b)^2 \left((\kappa - \kappa_b)^2 - 4 J^2\right)} - \frac{1}{\kappa^2} \right), \label{DE11} \\
\Delta E_{diss} & \equiv E^S_{B, \text{sup-opt}} - E_{\text{single-batt}} = \frac{4 F^2 \left( \frac{\Gamma^2 \kappa}{\kappa_b \left(2 \Gamma + \sqrt{\kappa} \sqrt{\kappa_b}\right)^2} - 1 \right)}{\kappa^2}. \label{DE12}
\end{align}
We plot these functions for typical parameter settings in Fig. \ref{DeltaE}. As shown, with sufficient isolation of the battery (i.e., sufficiently small $\kappa_b$), the super-optimal charging process is significantly more efficient than the direct charging scenario. In fact, in the limit of perfect isolation of the battery, the super-optimal charging allows for unbounded values of energy stored $\lim_{\kappa_b\rightarrow 0}\Delta E_{diss}=\infty$, while energy in the coherent setting is bounded: $\lim_{\kappa_b\rightarrow 0}\Delta E_{coh}=\frac{4F^2}{4 J ^2-\kappa^2}$. The charger-mediated setting with coherent coupling is beneficial only in the regime of extremely small $\kappa_b$. It is interesting to note that the stationary energy of single-battery setting (Eq. (\ref{DE10})) is the same as the stationary energy of the charger in the super-optimal charging (Eq. (\ref{EnergyA2opt})). Therefore, extension of the noisy single battery system via super-optimal setting effectively allows for creating a charged battery system with energy not bounded by the dissipation on the original charger. In turn, the stationary state of the charger is not affected by the extension.

We conclude the characterization of super-optimal charging with a remark about transient behaviour of this charging method. The effective coupling strength of the dissipative interaction depends entirely on the value of $\Gamma$ (see Eq. (\ref{ME3})). Therefore, while super-optimal scenario for more isolated batteries relies on reduced $\Gamma_{b}$ (as implied by reduced $\kappa_{b}$ via Eq. (\ref{x})), the limit $\lim_{\kappa_{b}\rightarrow 0}$ does not impose divergent times for reaching a given value of energy of the battery. Consequently, the advantage of super-optimal charging over the single-battery setting can be observed within the same time scales, and therefore offers significant practically-achievable boost in capacity of quantum batteries.

\textit{Summary and Outlook}.---In this work we investigate optimal strategies for charging quantum batteries by a laser coherent field in a charger-battery setting. We start from  optimisation of  external energy source (its frequency), by examining the built-in detunings of the charging setup. We then move to setups which introduce engineered shared reservoir into charging processes. Our investigation reveals that replacing the coherent interaction with a dissipative interaction via an engineered shared reservoir not only reproduces the same optimal charging outcomes but also allows for further optimization over the internal structure of the charging device (parameters of the coupling with the engineered reservoir), giving rise to super-optimal charging. As a result we observe that the total energy within the entire system attains a higher value in the steady state. Above all, we show that the ultimate advantage of a shared reservoir lies in its ability to facilitate a beneficial redistribution of energy within the charger-battery system, with a notable increase of energy in the desired location—namely the battery, presenting a significant advancement in quantum battery charging. This advantage is observable already in the transient regime. Therefore, we conclude that the super-optimal charging achieved through careful optimization of the engineered reservoir paves the way for transformative developments in high-performance energy solutions. 

\textit{Acknowledgements}.---BA acknowledges support from National Science Center, Poland within the QuantERA II Programme (No 2021/03/Y/ST2/00178, acronym ExTRaQT) that has received funding from the European Union’s Horizon 2020. BA also acknowledges Robert Alicki for helpful discussions. SB acknowledges funding by the Natural Sciences and Engineering Research Council of Canada (NSERC) through its Discovery Grant, funding and advisory support provided by Alberta Innovates through the Accelerating Innovations into CarE (AICE) -- Concepts Program, and support from Alberta Innovates and NSERC through Advance Grant.

\bibliography{References}

\clearpage
\onecolumngrid
\appendix

\section{Global View Analysis}\label{AppendixA}

To have a global understanding of the dynamics of the charging process, we rewrite the system Hamiltonian
\begin{equation}\label{Hamiltonian0}
H=\omega_Aa^\dagger a + \omega_Bb^\dagger b + (J a^\dagger b + J^* b^\dagger a) + F(e^{i\omega_L t}a + e^{-i\omega_L t}a^\dagger),
\end{equation}
in the properly selected basis, which depends on the value of $J$. We introduce super-mode operators
\begin{align*}
C_{+}=\frac{a+e^{i\phi}b}{\sqrt{2}},\numberthis\label{C+}\\
C_{-}=\frac{a-e^{i\phi}b}{\sqrt{2}},\label{C-}\numberthis
\end{align*}
with eigen-frequencies $\omega_{\pm}=\omega\pm |J|$, which satisfy commutation relations $[C_{+},C_{-}]=[a,b]=0$ and $[C_{\pm},C_{\pm}^{\dagger}]=[a,a^{\dagger}]=[b,b^{\dagger}]=1$. We make $C_{\pm}$ depend on the phase of the coupling $J=|J|e^{i\phi}$. From Eqs. \eqref{C+} and \eqref{C-} we have 
\begin{align*}
    a=\frac{C_{+}+C_{-}}{\sqrt{2}},\numberthis\label{a}\\
    b=e^{-i\phi}\frac{C_{+}-C_{-}}{\sqrt{2}},\numberthis\label{b}
\end{align*}
and hence we obtain
\begin{align*}
H&= \left(\Delta+|J|\right)C_{+}^{\dagger}C_{+}+\left(\Delta-|J|\right)C_{-}^{\dagger}C_{-} + \frac{F}{\sqrt{2}}\left(e^{i\omega_{L}t}C_{+}+e^{-i\omega_{L}t}C_{+}^{\dagger}\right) + \frac{F}{\sqrt{2}}\left(e^{i\omega_{L}t}C_{-}+e^{-i\omega_{L}t}C_{-}^{\dagger}\right).\numberthis
\end{align*}
At the same time, the master equation given in Eq. (\textcolor{blue}{3}) of the manuscript can be rewritten as 
\begin{equation}\label{ME2}
    \dot{\rho}=-i[H,\rho] + \Gamma\mathcal{D}_z[\rho] + \kappa_a\mathcal{D}_{z_{a}}[\rho]  + \kappa_b\mathcal{D}_{z_{b}}[\rho],
\end{equation}
where
\begin{equation}\label{z1}
z_a=:p_{a}^{a}C_{+}+p_{b}^{a}C_{-}, 
\end{equation} 
with $p_{a}^{a}=\frac{1}{\sqrt{2}}$, $p_{b}^{a}=\frac{1}{\sqrt{2}}$ based on Eq. \eqref{a}, 
\begin{equation}\label{z2}
z_b=:p_{a}^{b}C_{+}+p_{b}^{b}C_{-},
\end{equation} 
with $p_{a}^{b}=\frac{e^{-i\phi}}{\sqrt{2}}$, $p_{b}^{b}=-\frac{e^{-i\phi}}{\sqrt{2}}$ based on Eq. \eqref{b}, and
\begin{equation}
    z=p_{a}\frac{C_{+}+C_{-}}{\sqrt{2}}+p_{b}e^{-i\phi}\frac{C_{+}-C_{-}}{\sqrt{2}}=p_{a}^{z}C_{+}+p_{b}^{z}C_{-}.
\end{equation}
In the basis defined by $p_{a}^{z}=\frac{p_a+e^{-i\phi}p_{b}}{\sqrt{2}}$ and $p_{b}^{z}=\frac{p_a-e^{-i\phi}p_{b}}{\sqrt{2}}$, it is evident that local dissipators $\mathcal{D}_{a}$ and $\mathcal{D}_{b}$, expressed in the basis of $a$ and $b$, consistently translate into global dissipators $\mathcal{D}_{z_a}$ and $\mathcal{D}_{z_b}$ in the new basis. Conversely, by setting $p_{a}=\pm p_{b}e^{-i\phi}$, we ensure that the dissipator $\mathcal{D}_z$, which is global in the basis of $a$ and $b$, becomes local in the basis of $C{+}$ and $C_{-}$.

To formulate equations of motion in the basis of $C_{+}$ and $C_{-}$, we introduce $\mu_{z}=-p_{a}^{z}(p_{b}^{z})^{}$, $\mu_{a}=-p_{a}^{a}(p_{b}^{a})^{}=-\frac{1}{2}$, and $\mu_{b}=-p_{a}^{b}(p_{b}^{b})^{*}=\frac{1}{2}$. Subsequently, leveraging the linearity of the master equation and accounting for the adjusted amplitudes $F$ associated with both degrees of freedom, we arrive at:
\begin{align*}\label{G1}
    \frac{d\av{C_{+}}}{dt}= &-\frac{\Gamma|p_{a}^{z}|^2+\frac{\kappa_{a}+\kappa_{b}}{2}+2i(\Delta+|J|)}{2} \av{C_{+}} + \left(\frac{\Gamma}{2}\mu_{z}^{*}+\frac{\kappa_{b}-\kappa_{a}}{4}\right)\av{C_{-}}-i\frac{F}{\sqrt{2}},\numberthis
\end{align*}
\begin{align*}\label{G2}
    \frac{d\av{C_{-}}}{dt}=&-\frac{\Gamma|p_{b}^{z}|^2+\frac{\kappa_{a}+\kappa_{b}}{2}+2i(\Delta-|J|)}{2} \av{C_{-}} + \left(\frac{\Gamma}{2}\mu_{z}+\frac{\kappa_{b}-\kappa_{a}}{4}\right)\av{C_{+}}-i\frac{F}{\sqrt{2}},\numberthis
\end{align*}
\begin{align*}\label{G3}
    \frac{d\av{\dg{C_{+}}C_{+}}}{dt}=&-\Big(\Gamma|p_{a}^{z}|^2+\frac{\kappa_{a}+\kappa_{b}}{2}\Big)\av{\dg{C_{+}}C_{+}} + 2\Re\Big\{\Big(\frac{\Gamma}{2}\mu_z^{*}+\frac{\kappa_{b}-\kappa_{a}}{4}\Big)\av{\dg{C_{+}}C_{-}}\Big\} - 2\Im\Big\{\frac{F}{\sqrt{2}}\av{C_{+}}\Big\},\numberthis
\end{align*}
\begin{align*}\label{G4}
   \frac{d\av{\dg{C_{-}}C_{-}}}{dt}=&-\Big(\Gamma|p_{b}^{z}|^2+\frac{\kappa_{a}+\kappa_{b}}{2}\Big)\av{\dg{C_{-}}C_{-}} + 2\Re\Big\{\Big(\frac{\Gamma}{2}\mu_z^{*}+\frac{\kappa_{b}-\kappa_{a}}{4}\Big)\av{\dg{C_{+}}C_{-}}\Big\} - 2\Im\Big\{\frac{F}{\sqrt{2}}\av{C_{-}}\Big\},\numberthis
\end{align*}
\begin{align*}\label{G5}
   \frac{d\av{\dg{C_{+}}C_{-}}}{dt}=&-\frac{\Gamma(|p_{a}^{z}|^2+|p_{b}^{z}|^2)+\kappa_{a}+\kappa_{b}-4i|J|}{2}\av{\dg{C_{+}}C_{-}} + \Big(\frac{\Gamma}{2}\mu_z+\frac{\kappa_{b}-\kappa_{a}}{4}\Big)\av{\dg{C_{+}}C_{+}}  + \Big(\frac{\Gamma}{2}\mu_z+\frac{\kappa_{b}-\kappa_{a}}{4}\Big)\av{\dg{C_{-}}C_{-}}\\\nonumber
    &-i\frac{F}{\sqrt{2}}\av{C_{+}}+i\frac{F}{\sqrt{2}}\av{C_{-}}.\numberthis
\end{align*}
By selecting $\phi=0$ the expressions for $p_{a}^{z}$ and $p_{b}^{z}$ become $p_{a}^{z}=\frac{p_a+p_{b}}{\sqrt{2}}$ and $p_{b}^{z}=\frac{p_a-p_{b}}{\sqrt{2}}$. Specifically, when $p_a=p_b$, we obtain $p_{a}^{z}=p_a$ and $p_{b}^{z}=0$, resulting in $\mu_z=0$. Referring to Eq. \eqref{G4}, this implies that the global mode $C_-$ experiences a decoherence-free evolution concerning the shared reservoir, mimicking a scenario with no shared reservoir. And choosing $\phi=\pi$ it is the global mode $C_+$ that experiences a decoherence-free evolution concerning the shared reservoir.

However, for $\phi\neq0$, the correction to the built-in detuning is more complicated due to the presence of the shared reservoir: 
\begin{align*}\label{Deltaopt}
    \Delta_{opt} = -\frac{6^{1/3} \left(8\Delta^2_{in} - 2 \Gamma^2 - \Gamma_a^2 - \Gamma_b^2 - 2 \Gamma_a \kappa_a - \kappa_a^2 - 2 \Gamma_b \kappa_b - \kappa_b^2 \right) - (K - M)^{2/3}}{2\times6^{2/3}(K - M)^{1/3}},\numberthis
\end{align*}
where
\begin{equation}\label{K}
    K =: 36|J|\Gamma\cos{\phi} \left(\Gamma_a + \Gamma_b + \kappa_a + \kappa_b\right),
\end{equation}
\begin{equation}\label{M}
    M =: 6^{1/2}\left(\left(8\Delta^2_{in} - 2 \Gamma^2 - \Gamma_a^2 - \Gamma_b^2 - 2 \Gamma_a \kappa_a - \kappa_a^2 - 2 \Gamma_b \kappa_b - \kappa_b^2 \right)^3 - 216\Delta^2_{in} \Gamma^2 \left( \Gamma_a - \Gamma_b - \kappa_a - \kappa_b \right)^2 \cos{\phi}^2\right)^{1/2}.
\end{equation}
Putting $\Gamma=0$ we obtain the optimal detuning given in Eq. (\textcolor{blue}{13}) of the main text.

\section{Total Energy Comparison: Before and After Optimization}\label{AppendixB}

\begin{figure}[t]
    \centering
    \includegraphics[width=0.5\textwidth]{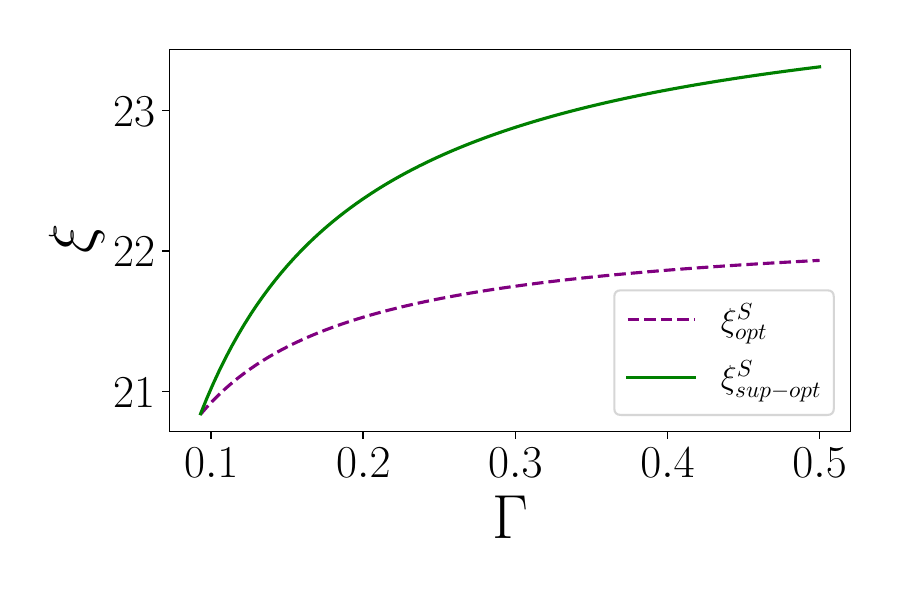}
    \caption{$\xi^S$ and $\xi^S_{opt}$ versus $\Gamma$. For $\xi^S$ we have the optimized $p_a=p_b=1$ and for $\xi^S_{opt}$ we have $p_{a}=(\kappa_a/\kappa_b)^{1/4}$, $p_{b}=(\kappa_a/\kappa_b)^{-1/4}$. In both cases we have $\Delta_{opt}=0$, $\kappa_a=0.05$, and $\kappa_b=0.01$.}
    \label{XiComparison}
\end{figure}

Replacing the coherent interaction $J$ by the shared reservoir with dissipation rate $\Gamma$ and then performing the optimization A significant change in both the overall and local energies of the system is observed. In Fig. \ref{XiComparison} the total energy of the system $\xi^S$, in the steady state, before (blue line) and after (red line) the optimization are illustrated. As is seen a significant increase is observed due to the optimization. Therefore, optimization preserves more energy, in the charger-battery system, against dissipation.

\end{document}